\documentclass[conference]{IEEEtran}
\IEEEoverridecommandlockouts
\usepackage{url}
\usepackage{soul,color}

\usepackage{cite}
\usepackage{amsmath,amssymb,amsfonts}
\usepackage{algorithmic}
\usepackage{graphicx}
\usepackage{textcomp}
\usepackage{xcolor}
\def\BibTeX{{\rm B\kern-.05em{\sc i\kern-.025em b}\kern-.08em
    T\kern-.1667em\lower.7ex\hbox{E}\kern-.125emX}}
\begin{document}

\title{Life cycle management of automotive data functions in MEC infrastructures 
}

\makeatletter
\newcommand{\linebreakand}{%
  \end{@IEEEauthorhalign}
  \hfill\mbox{}\par
  \mbox{}\hfill\begin{@IEEEauthorhalign}
}
\makeatother

\author{\IEEEauthorblockN{Mikel Seron, Angel Martin}
\IEEEauthorblockA{\textit{Department of Digital Media \& Communications} \\
\textit{Vicomtech Foundation}\\
Basque Research and Technology Alliance\\
San Sebasti\'an, 20009 Spain\\
Email: mseron@vicomtech.org, amartin@vicomtech.org}
\and
\IEEEauthorblockN{Gorka Velez}
\IEEEauthorblockA{\textit{Department of Intelligent Transportation Systems \& Engineering} \\
\textit{Vicomtech Foundation}\\
Basque Research and Technology Alliance\\
San Sebasti\'an, 20009 Spain\\
Email: gvelez@vicomtech.org}
\linebreakand 
}

\maketitle

 \IEEEoverridecommandlockouts
 \IEEEpubid{\begin{minipage}{\textwidth}\ \\\\\\\\\\[12pt]
\textbf{M. Seron, A. Martin and G. Velez, "Life cycle management of automotive data functions in MEC infrastructures," 2022 IEEE Future Networks World Forum (FNWF), Montreal, QC, Canada, 2022, pp. 407-412, doi: 10.1109/FNWF55208.2022.00078. \copyright 2022 IEEE.  Personal use of this material is permitted.  Permission from IEEE must be obtained for all other uses, in any current or future media, including reprinting/republishing this material for advertising or promotional purposes, creating new collective works, for resale or redistribution to servers or lists, or reuse of any copyrighted component of this work in other works.}
 \end{minipage}}
 
\begin{abstract}
Cars capture and generate huge volumes of data in real-time, including the driving dynamics, the environment, and the driver and passengers’ activities. With the proliferation of Connected and Automated Mobility (CAM) applications, the value of vehicle data is getting higher for the automotive industry as it is not limited to onboard systems and services. This paper proposes an architecture that exploits Multi-access Edge Computing (MEC) technology of 5G networks to enable data monetisation. It employs a virtualisation framework that instantiates on consumer demand pipelines that process data samples according to Service Level Agreement (SLA) policies, licensing terms and Region Of Interest (ROI) clusters with a privacy-centric design. In addition, the aspects that need to be considered when creating a data marketplace for the automotive sector are identified while highlighting the design features that go beyond the current scientific and market solutions.


\end{abstract}


\section{Introduction}
Vehicle manufacturers are adding advanced and sophisticated sensors to onboard Advanced Driver Assistance Systems (ADAS) to enable them to understand the environment better, supporting the drivers to enforce safety or perform complex manoeuvres. Hence, Original Equipment Manufacturers (OEMs) compete to incorporate cutting-edge ADAS features. Furthermore, manufacturers include connectivity devices to enable telemetry and some innovative services originally related to maintenance diagnostics and fidelity. Innovative Connected and Automated Mobility (CAM) applications mainly focus on cooperation and coordination of vehicular systems exploiting vehicle sensoring and connectivity features \cite{velez2020}. 

However, vehicle manufacturers are aware that their business horizons are not just related to driving-aid and support. OEMs and Tier-1s clearly identify that a significant part of their future income will come from exploiting the data captured and produced in the vehicle environment\cite{mckinsey2021}. Thus, Internet giants such as Google, Apple, and Amazon are trying to position themselves as key actors in the automotive data marketplace.

Currently, data generated in the cars are data silos processed by onboard systems or feeding cloud systems from the manufacturer, but in any case, only the manufacturer can access or process them. OEMs and Tier-1 suppliers are under pressure to develop their own data-driven value propositions by  leveraging digital ecosystems and enforcing cooperation to fully influence  data-driven business opportunities beyond autonomous driving and connected mobility domains, not only to improve existing processes and functions. 

In order to favour the creation of an automotive data ecosystem, it is necessary to provide common data pipelines which connect data producers and consumers. However, data privacy and ownership are core topics to consider when processing automotive data. Here, Multi-access Edge Computing (MEC) infrastructures can benefit from their close position to the data producers to process data as soon as they are transmitted towards the platform employed by consumers to access data. Furthermore, the capillarity of the MEC architectures fosters distributed processing and local scalability in contrast to cloud infrastructures.

The platform architecture produced in the context of the Horizon 2020 project 5GMETA and introduced in this paper considers different aspects, including the mandatory privacy processing to remove any personal data, the tagging and control of licensing models to keep data producer ownership and grant access to applications satisfying the applicable terms, the generation of data flows anchored to virtualised geographical areas, targeted by particular applications or users, and the management of virtualised pipelines to foster efficient processing of data samples in the MEC. It also includes different solutions to host Digital Twin-based services in the MEC for those applications requiring low latency. The platform includes mechanisms to account for the volume of data consumed and the computing assets employed in the MEC for the automatically deployed pipelines.

This paper is structured as follows. First, section \ref{sec:related} focuses on requirements for future applications based on automotive data and virtualisation solutions to process data efficiently. Then, section \ref{sec:solution} presents the implemented architecture providing a cloud platform which indexes and operates different MEC infrastructures where data pipelines are instantiated and run on demand to process the produced data targeted by data consumers. Finally, section \ref{sec:conclusions} summarises the paper topics and outlines some open questions.



\section{MEC for data processing}
\label{sec:related}
This section analyses the requirements to be considered by platforms executing data pipelines for the automotive sector and solutions to host in the MEC virtualised containers running data pipelines.

\subsection{Data pipelines for the CAM sector}


The different data functions that an automotive data platform needs to provide are listed below:

\begin{itemize}
    \item Interoperability. A vehicle captures heterogeneous data such as vehicle attributes, status, position, and dynamics, external sensors such as rain or temperature, cameras or Light Detection and Ranging (LIDAR), driver status, or online data. Here, different formats are employed including text, video and point clouds, as indicated by ETSI \cite{etsi2009}, 3GPP \cite{3gpp2020}, 5GAA \cite{5gaa2020} and CAR 2 CAR Communication Consortium \cite{c2c2019}. Some are open, while others are proprietary with binary formats and packaging. Therefore, it is evident that interoperability is a major challenge. Moreover,  a time synchronisation method such as Network Time Protocol (NTP) is needed to correlate different data sources with different vehicle clocks,
    \item Privacy. The sensors can capture and transmit the information which should be removed before being processed or shared. In this regard, the MEC has a perfect position, as close to the user as possible, to preserve privacy as soon as the data reaches the network \cite{soua2018}. Here, some blacklist fields can be removed from JavaScript Object Notation (JSON) formatted messages, and vehicular plates or faces can be blurred on camera images.
    \item Service Level Agreement (SLA). It is important that the resource allocation in the MEC to process data samples considers the pricing plan contracted and the complexity of data type to process by the pipeline \cite{belogaev2020}. The SLA is also translated into the sampling rate applied to the data to provide sub-sampled data flows, meaning lower data velocity and volumes to primary consumers.
    \item Filtering. In automotive data, geoposition has a significant presence. The CAM applications use them to better match application responses to the local context of produced data \cite{velez2020}. Furthermore, the declaration of the applicable license model by the data producer as owner limits the access to data to those consumers who satisfy the terms and conditions. Thus, the configuration of target data types, eligible license models, and Regions of Interests (ROIs) and the production of data channels with the applied filters are essential for CAM applications to focus on valid and relevant data processed by pipelines.
\end{itemize}


It becomes evident that the operation of the life cycle of virtualised pipelines in the MEC is crucial to dynamically respond to data consumption requests. The automated operation of pipeline instantiation makes the difference in efficiently responding to demand. Furthermore, the ability to reuse pipelines processing a specific data type for common or heterogeneous SLAs and licenses can significantly improve the efficiency at the MEC.


The automotive industry is exploring technologies which are already familiar to them in the production value chain, including Internet of Things (IoT) messaging technologies such as Advanced Message Queuing Protocol (AMQP), Message Queuing Telemetry Transport (MQTT), or Apache Kafka, feeding their Supervisory Control And Data Acquisition (SCADA) systems also valid for telemetry and real-time data export outside the vehicle \cite{lazidis2022}. Some platforms to index all the data sources and manage the records are GreenGrass (probing driver), DynamoDB (persistency) and RTOS (processing) from Amazon, Beam (container-based pipeline programming), and Flink (live processing) from Apache, Kuksa (inventory and persistency) from Eclipse and Vehicle Management Platform from Bosch. However, in terms of privacy, they provide access to data producer endpoints and, in terms of efficiency, they do not scale/provision pipelines based on consumer demand but on production volume; in terms of ownership, they do not deal with licenses as the producers and consumers are intended to be the same or under the same managers or operators. Here is where our approach makes the difference.

\subsection{VNFs for data pipelines}



The MEC platforms should aid in delivering real-time data pipelines for car-captured and generated data through data privacy, interoperability, computing, and security functions embodied in Virtual Network Functions (VNFs) deployed in a Network Functions Virtualization (NFV) enabled architecture.

Prior to the introduction of NFV, it was common practice to deploy network applications and services using specialised proprietary hardware and software that could only be used in particular installations, being an unyielding system. NFV overcomes challenges like reducing capital and operating expenses and satisfying the growing demand for mobile services. Due to NFV, software and services may be deployed in any environment, enabling them to be virtualised. NFV Management and Orchestration (NFV MANO) is a crucial infrastructure for unlocking the full potential of the virtualisation of network functions. The most relevant implementations are Open Source MANO (OSM), hosted by ETSI, and Open Network Automation Platform (ONAP), supported by Linux Foundation.

Furthermore, microservices can be used to build VNFs. Microservice architecture is a distinctive approach to developing lightweight software that builds single-function modules with well-defined interfaces and operations. This architecture enables the rapid, frequent, and reliable delivery of large, complex applications in a modular way. Microservices can run small software components that share an Operating System (OS). This is known as OS virtualisation, a similar approach to the traditional virtualisation technology, where a physical server is split into smaller components as Virtual Machines (VMs). However, microservices are more efficient than VMs because they share an OS, whereas each VM requires its own OS, consuming more resources. Like Microservices, containers have also been gaining popularity as an indispensable ingredient to flexible architecture. A container is a bundling of an application and all its dependencies as a package that allows it to be deployed quickly and consistently regardless of the environment. Containers and microservices include specific approaches such as Docker and Linux containers, which have been mainly popularised.

To fully operate a container-based microservices architecture, an orchestration tool is needed. An orchestrator automates containers' deployment, management, scaling, and networking. Managing the lifecycle of containers is essential. Container orchestration tools provide a framework for managing containers and microservices architecture at scale. It also grants the scalability of every service. Many container orchestration tools can be used for container lifecycle management, such as Kubernetes and Docker Swarm.

The automated operation needs monitoring to trigger programmed actuation policies and rules. Thus, the infrastructures need to be monitored, given the importance of preventing errors and system failures. In addition, obtaining metrics and logs of the deployed applications is key to measuring their performance, status, or speed in real-time and automatically reacting to a given event. Multiple monitoring and visualization tool options are available for container orchestrators, such as Prometheus/Grafana and Elasticsearch/Logstash/Kibana (ELK) stacks. Monitoring the framework and its workloads allows scheduling the servers' resources, reserving a portion of the Central Processing Unit (CPU) and memory resources for using the data pipelines, and fulfilling the requested SLA.

These sophisticated technologies can also be applied to the MEC standard introduced by ETSI, to minimise the latency when serving data by pushing computing and storage closer to the data source. Thanks to MEC, the amount of raw data traffic traversing the core network is decreased, thus also relieving the pressure on the remote cloud infrastructure. Additionally, CAM must imply process-intensive and low-latency computing and communication services. Due to these factors, cloud connections are frequently insufficient. Remote data processing and storage are unreliable and slow for many time-critical CAM services, making MEC a suitable solution.

The deployment of microservices at the network edge as VNFs provides a highly versatile and flexible framework for pipeline deployment while fostering real-time data processing with high bandwidth connectivity and extremely low latency.

\subsection{MEC for automotive data}
The application of MEC technology to boost the response of CAM services or to empower the processing of automotive data is a novel research field with some works studying edge computing in vehicular communications.

A MEC architecture and some Application Programming Interfaces (APIs) for C-V2X systems are proposed in \cite{wang2019}, checking the feasibility of hosting CAM use cases in the MEC. The work presented in \cite{campolo2019} goes a step further in describing a Docker-based ETSI-compliant MEC platform, taking benefit of the microservice paradigm. Thanks to the modularity of the microservices, a service migration procedure is proposed.

In regard to mobility inherent to users of cellular networks, the handovers between different MEC servers have gained a lot of attention. The problem of maintaining service continuity and synchronisation of relevant data among multiple MEC servers to support vehicular applications is studied in \cite{ojanpera2018} and an architecture is proposed for this end.

Two different frameworks for integrating big data analytics with vehicular edge computing are proposed in \cite{zhang2018} and \cite{zhou2018}. These frameworks select the offloading infrastructures from the available ones satisfying heterogeneous requirements. Another approach involves a hierarchical model for resource management at MECs targeting latency and energy efficiency requirements \cite{duan2022}.

Two topics are paramount when studying the fields related to vehicular data sharing using edge computing platforms and cellular networks: privacy and efficiency. Regarding privacy concerns, secure peer-to-peer data sharing systems are necessary to prevent second-hand data sharing without authorisation, as the one based on blockchain proposed in \cite{kang2019}. Regarding processing and communications efficiency, a hierarchical edge framework performing computation offloading and content caching is proposed \cite{qin2020} to minimise network communication overheads for recalled tasks.

All the presented works ignore critical aspects of vehicle data platforms to connect data sensors with future CAM applications. They focus on processing all the produced data, lacking a mechanism to trigger data processing offloading in the MEC when the consumption demand is there. All the data processed and not consumed mean a big overhead and an inefficient approach. Additionally, they put SLA levels aside ignoring business models when allocating computing resources. 

\section{MEC for Data-centric CAM Applications}
\label{sec:solution}

The following section describes the implemented architecture of the data processing platform developed inside the Horizon 2020 project 5GMETA,  funded by the EU\cite{5gmeta}.

\subsection{Cost-effective MEC for data pipelines}

Our approach designs a platform which: i) performs data processing in the MEC based on consumer demand; ii) enables filtering on geographical and data types criteria to get just relevant data; and iii) applies SLA levels to different asset allocation profiles.

As depicted in Figure \ref{fig:workflow}, the described framework is an open data-centric IoT live messaging platform for CAM services and applications where the security, privacy, scalability, interoperability, and licensing features are provided by the 5G networks functions executed at the edge to gain zero latency, capillarity, and geo-driven networking. The reference architecture embodies four layers:

\begin{itemize}
    \item Sensors and Devices (e.g., Light Detection and Ranging (LIDARs), cameras). It is used to generate road data, whether mounted on vehicles or different Road Side Units (RSUs) in the vehicles' surroundings. All of them access to the MEC infrastructure through a 5G connection.
    \item 5G Network with MEC infrastructure. Represents the main 5G core functions, 5G New Radio, and the platform's MEC system. This layer is mainly based on the 5G features that allow data to be routed with minimal latency to the edge platform. It also provides a virtualisation Infrastructure connected to the Base Station of the 5G infrastructure. This layer is directly connected to the cloud platform.
    \item Cloud Platform. This layer addresses the data management and monetisation aspects and requirements, allowing users to subscribe to different licensing types. It also bundles third-party APIs to route the data the users have subscribed to according to a pre-defined SLA type. It has a record of all the registered MECs.
    \item Third-party services. Where the applications that have authorised access to the data sit.
\end{itemize}

\begin{figure}[htbp]
\centerline{\includegraphics[width=0.95\columnwidth]{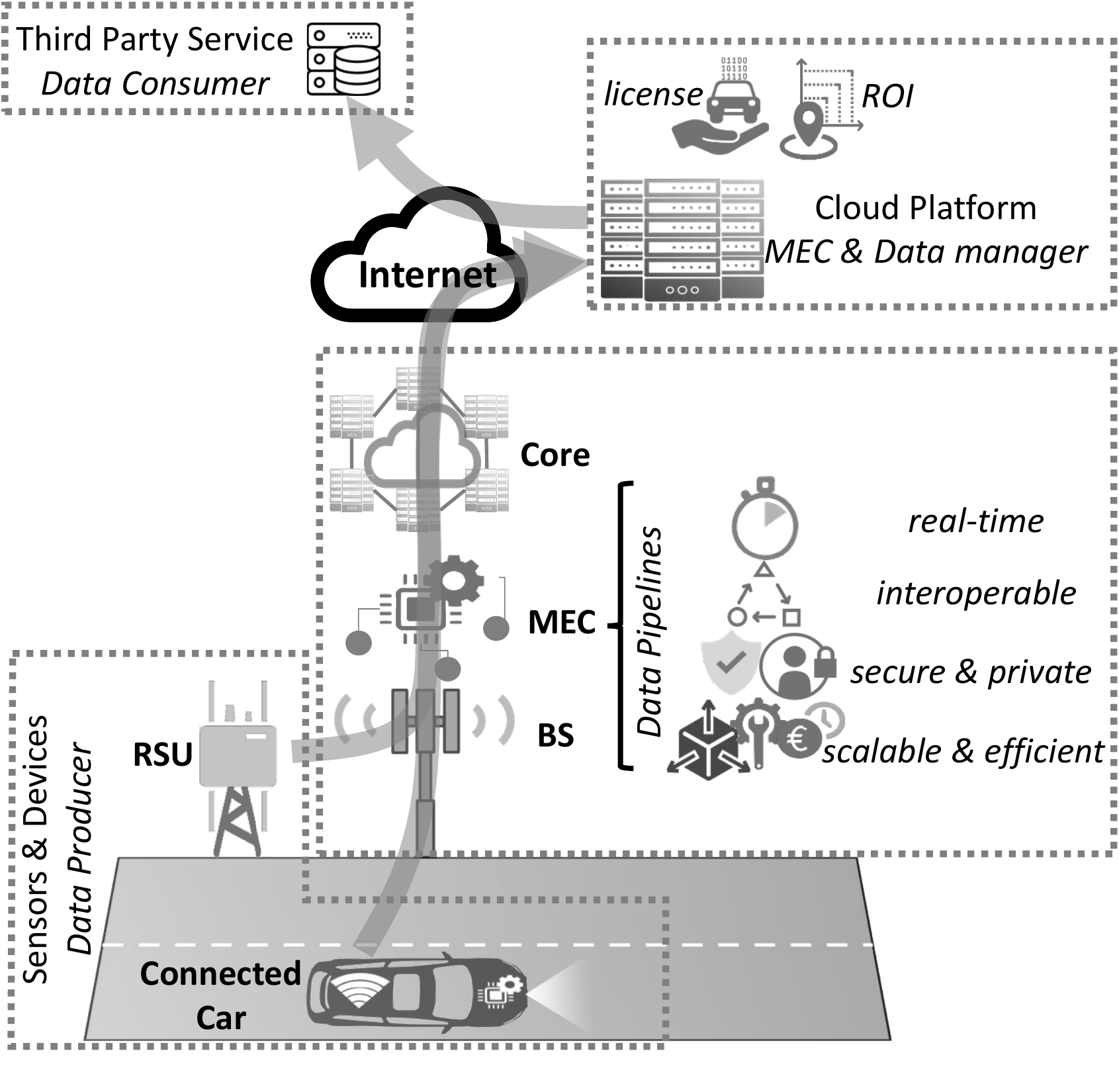}}
\caption{Automotive Data Workflow.}
\label{fig:workflow}
\end{figure}

Concerning the data workflows and permitted directions, both upload and download directions are considered.
\begin{itemize}
    \item Upload: This is the main data flow delivering data from sensors, vehicles, and RSUs to CAM services. When a CAM application selects a data type and a ROI, the containers of a pipeline processing the specific data type are deployed in the selected MECs. The data is only processed in the MEC and forwarded to the Cloud when a CAM service has selected a specific data type from a serving MEC.
    \item Download: Spontaneous, discrete, and lightweight alarms and notifications sent as broadcast messages from CAM applications to Sensors and Devices subscribed to the notification data type. This means no need for a specific function or container to deliver download messages.
\end{itemize}
So, the upload direction provides continuous live data feeds, while the latter is mainly designed to broadcast notifications and updates to all subscribed systems in a MEC.

The access level of the third-party application to the data available within the platform may differ depending on the application requirements. 
The consumed data volumes are accounted for in the cloud platform, and the required computing assets allocated for the pipelines process raw data from sensors and devices at the edge and forward the resulting data to the cloud.

However, if a specific application has strict requirements when it comes to latency, data can be directly reached from the MEC platform in real time. In this case, the consumed data is charged to the application based on the volume of produced data declared by the MEC service in the cloud infrastructure.

Another aspect of the platform is the possibility of hosting CAM services in the MEC, such as Digital Twin-based applications requiring real-time processing and actuation.

The deployment of pipelines in the MEC is triggered by data consumers (i) interested in processing and consuming a specific data type from a particular location (ii). When a CAM application selects a data type and if the SLA of the CAM application supports the required resources, the containers of a pipeline processing the specific data type are deployed in the corresponding MEC(s) (iii). Access to the data will also be limited to match the license terms defined by data owners. The data is only processed in the MEC and forwarded to the Cloud when a CAM service has selected a specific data type from a serving MEC; otherwise, data pushed to the MEC is immediately discarded.

\subsection{MEC architecture for automotive data}

The purpose of this section is to describe an NFV-enabled architecture for deploying data pipelines embodied in VNFs for car-generated data. This architecture is depicted in Figure \ref{architecture}.

\begin{figure}[htbp]
\centerline{\includegraphics[width=\columnwidth]{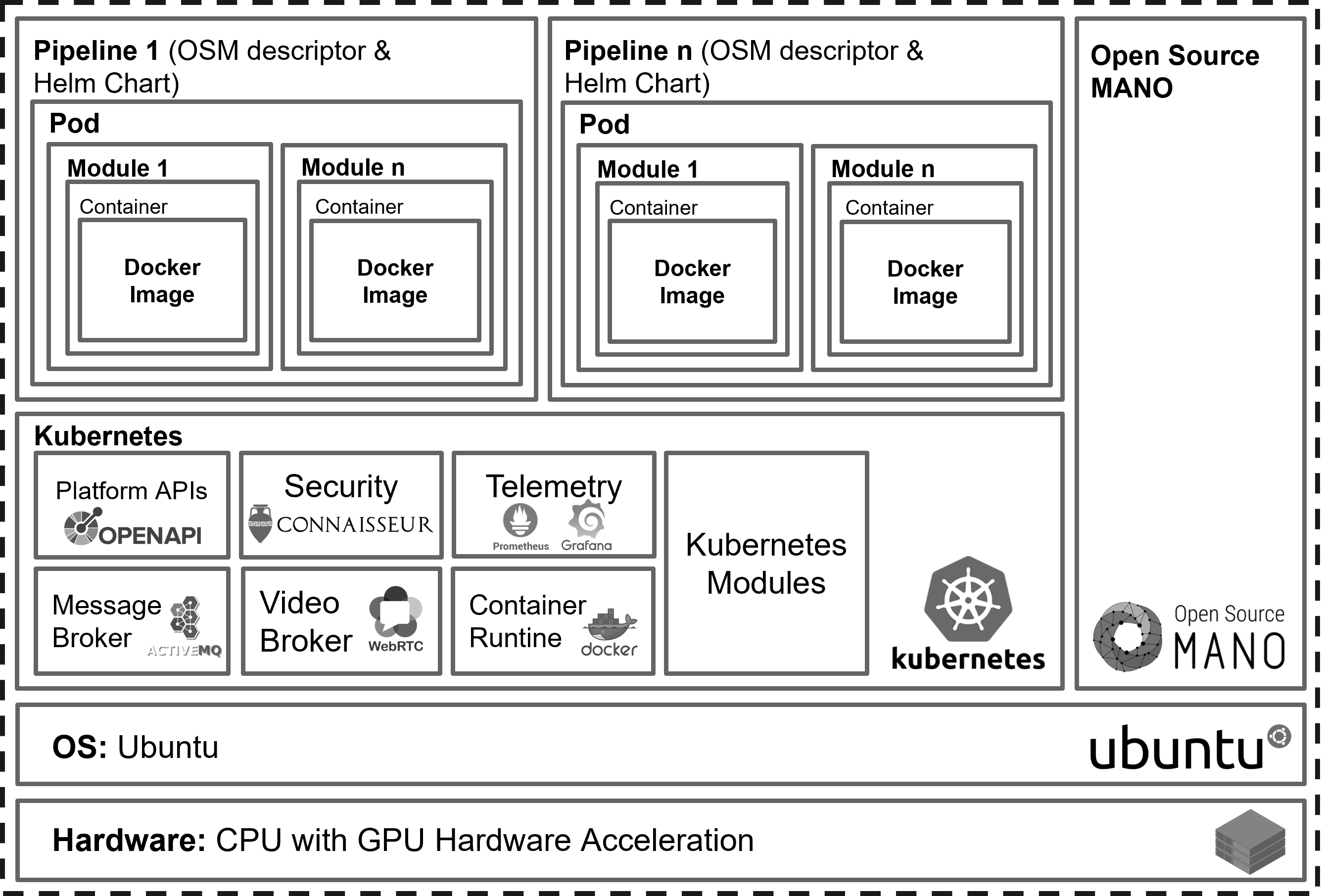}}
\caption{Architecture of the MEC platform.}
\label{architecture}
\end{figure}

First, concerning the life cycle management of data functions in the MEC, different industry solutions build the platform stack.

Kubernetes and Docker as container orchestrator and runtime, respectively, and OSM as orchestration framework are used to deploy the edge VNFs. Kubernetes is an open-source Container-as-a-Service platform (CaaS) for managing containerised workloads and services.  It handles scheduling onto nodes in a compute cluster, actively manages workloads, and makes managing and deploying containerised apps easy. Kubernetes supports horizontal and vertical auto-scaling, permitting the scale of the resources for data pipelines always according to the selected SLA level.

OSM, the management and orchestration software stack that follows the standards defined for NFV technology, carries out the task of orchestrating the specific data pipelines. It is in charge of ordering the actions related to the life cycles of the NFV services.

Both Kubernetes and Docker are industry standards widely employed in scientific and commercial setups. They are also compatible with commercial Infrastructure as a Service (IaaS) platforms such as Amazon Elastic Kubernetes Service (EKS). Thus, common virtualization technologies can also be used across the platform when managing a distributed MEC infrastructure using a central manager at the cloud. Furthermore, they are already supported by OSM as a prominent choice by 5GPPP projects to orchestrate platforms \cite{5gpppedge}. 

Kube-prometheus is used for platform monitoring, which is a pre-configured solution to collect metrics from all Kubernetes components. It includes a collection of Kubernetes manifests, Grafana dashboards, and Prometheus rules combined with documentation and scripts to provide easy-to-operate end-to-end Kubernetes cluster monitoring with Prometheus using Prometheus and Grafana tools.

Then, the life cycle manager is employed to instantiate or free data processing pipelines automatically. Every pipeline is wrapped into a VNF. The name is tied to the data type to be processed, meaning a specific topic in the message broker to be subscribed to, consuming data being processed, and a specific topic in the message broker where to publish processed data.

All the container images used by the different VNFs are hosted in a commercial cloud repository, such as DockerHub, accessible from any networked infrastructure. For ensuring the integrity and provenance of container images in the MEC platform, Connaisseur is used. This admission controller verifies every image signature and trust pinning before deploying any workload.

The data pipelines consume raw data samples from IoT frameworks. ActiveMQ is the central IoT messaging technology in the platform, an open-source, multi-protocol message broker. It has universal support from existing Sensors and IoT frameworks; it is a lightweight protocol and supports scalability for hierarchical data aggregation. The data producers push data into the MEC through AMPQ or MQTT protocol. The message broker exchanges data between the sensor and devices and the cloud for upload and download dataflow streams. For video data, the platform employs Web Real-Time Communication (WebRTC) for sending video as identified by Release 18 of 3GPP for real-time communications \cite{webrtc18}.

For handling ROI and geopositions, Microsoft's Bing Maps Tile System, where each region is represented by a single tile of the same shape and size, and geographical indexing with quadkeys is used \cite{tile}. A quadkey number, a one-dimensional array that combines zoom level, column, and row information, uniquely identifies a single tile position. The tiles and quadkey filtering permit data consumers to browse and quickly filter the locations where data is being produced. Every MEC also registers the tile where it is located and serves to data producers so that they can find the serving MEC infrastructure and do a handover across serving MECs as the vehicle moves.

Another key aspect is the design of common and consistent interfaces. The APIs are implemented to follow the specification of OpenAPIs 3.0, a language-agnostic interface to Representational State Transfer (REST) APIs. It satisfies vital aspects such as quick integration, documentation, access control, and web-based testing. Open Authorization (OAuth) 2.0 is also implemented as an authorisation protocol giving APIs limited access to user data. The authorisation is done in the cloud layer.

Once all the technology pieces have been described, the orchestration procedure is described below.

The deployment of pipelines in the MEC is triggered by data consumers interested in processing and consuming a specific data type. A third-party browses and requests from the available locations the dataflows available for specific relevant datatypes if data is being pushed from sensors and devices. The request is forwarded to the selected MEC infrastructure, where the brokers receive the data, and the pipeline processing starts. The orchestrator allocates resources according to the contracted SLA level. After checking assets availability, the target datatype's pipeline is deployed after downloading the respective containers from a repository and checking the signatures' validity. Once the pipeline is deployed, the data received by the broker and processed by the pipeline are delivered to the consumer through the cloud.

Each data type would need different computing resources to cope with incoming traffic. The number of assigned resources for a pipeline depends on the applicable SLA. For example, a GPU-enabled SLA is available through GPU virtualisation in the MEC for processing and optimising capabilities to process live video. GPU virtualisation is highly valuable for third-party CAM applications requiring low latency and stored in the MEC running heavy tasks like Computer Vision-based ones.

Following the same approach for the pipelines, the containers, in this case coming from the third parties for being deployed into the MEC, need further verification to be trusted and deployed inside the infrastructure. Two different aspects are then verified. First, if the declared data types are the only topics to which the container is subscribed. Second, if the volumes of synthetic data pushed into the subscribed topics are accurately declared. The validation of both tests provides a certain level of trust in the service before it is deployed in a specific MEC infrastructure. If any of these two steps are not satisfied, the container is banned and not deployed in any other MEC infrastructure.


\section{Conclusions and Future Work}
\label{sec:conclusions}
The CAM applications will benefit from the performance, reliability, and capacity promised by 5G cellular networks. Furthermore, the connected car data is expected to unlock new business for traditional actors and new entrants on top of innovative CAM applications. Here, the MEC architectures from 5G networks can fuel the creation of a common data marketplace where data owners produce and share their data and data services consume the available data demanding the processing of common pipelines in the MEC which deal with interoperability, privacy, and efficiency. This paper proposes an architectural approach designed and implemented to facilitate the custody of data ownership and filter data from specific ROIs. Furthermore, some novel features provided by the proposed architecture, when compared to market solutions, are the separation of data producers and consumers through dedicated IoT topics where pipelines read raw samples and write processed data, the allocation of computing resources adapted to the data consumption demand and the hosting of services which require low latency in MEC infrastructures applying the container-based life cycle management for pipelines to third-party services such as Digital Twin based applications.

This platform is being deployed and operated in three different pilot sites involved in the Horizon 2020 project 5GMETA. Thus, the proposed MEC infrastructure is introduced and tested in the French (Satory), Italian (Torino) and Spanish (San Sebastian) pilot sites where the servers are accessible via the 5G radio technology allowing the UE devices to access their services with very low latency.

In terms of efficiency, there is a lot of work to do to get the most from the computing resources used by the running pipelines, reusing processed data for different applications with different sampling rates from different SLAs, with partially overlapped ROIs and common eligible licenses while preventing promiscuous access to data.

Furthermore, the application of Network Slicing technologies to the proposed architecture and data workflows will isolate different data flows for specific applications. Specifically, applying RAN slicing techniques is essential to prioritise data communications for safety-related CAM applications over non-real-time-sensitive ones.



\section*{Acknowledgment}

This research was supported by the European Union’s Horizon 2020 research and innovation program under grant agreement No. 957360 (5GMETA project) and the Spanish Centre for the Development of Industrial Technology (CDTI), and the Ministry of Economy, Industry, and Competitiveness under grant/project \textit{CER-20191015 / Open, Virtualized Technology Demonstrators for Smart Networks (Open-VERSO)}.






\vspace{12pt}

\end{document}